\documentclass[final,5p,times,twocolumn,numbers]{elsarticle}
\usepackage{amssymb}
\usepackage{lipsum}
\usepackage{lineno,hyperref}
\usepackage{lineno}
\usepackage{graphicx}
\usepackage{subcaption}
\usepackage{amsmath}
\usepackage{verbatim}
\usepackage{color}
\usepackage{soul}
\usepackage[dvipsnames]{xcolor}
\usepackage{mathrsfs}
\modulolinenumbers[5]

\journal{Physics Letters B}

\begin{document}

\begin{frontmatter}

\title{First Law for Nonsingular Black Holes in 2D Dilaton Gravity}

\author[]{Peng Yu}
\author[]{Yuan Zhong\corref{cor1}}
\ead{zhongy@xjtu.edu.cn}
\affiliation{organization={MOE Key Laboratory for Nonequilibrium Synthesis and Modulation of Condensed Matter, School of Physics, Xi'an Jiaotong University},
            addressline={Xi'an 710049}, 
            city={Xi'an},
            postcode={710049}, 
            state={Shannxi},
            country={China}}
\cortext[cor1]{Corresponding author.}
\begin{abstract}
A central issue in the thermodynamics of nonsingular black holes is the apparent violation of the first law. In this work, we use 2D dilaton gravity as a simple theoretical setting to study this issue. We systematically construct a broad class of nonsingular black hole solutions with metric function $A(x)=f(x)+c$, through a procedure that is considerably simpler than in higher-dimensional theories. Using the Iyer-Wald covariant phase space formalism, we derive the correct energy formula and establish a consistent first law for this entire class of solutions. The apparent violation of the first law in a previous work arises because the energy used therein is not the Hamiltonian conjugate to the fixed time-translation generator adopted. Under the boundary conditions and normalization convention specified in the present work, the Iyer-Wald Hamiltonian energy is $E=-\frac{c}{2}$, and the first law is restored. Moreover, the energy formula agrees with the Casimir function in 2D dilaton gravity, thus confirming its interpretation as the physical black hole energy. Our results clarify the correct first law for 2D nonsingular black holes and may provide insights into the first law of nonsingular black holes in higher dimensions.
\end{abstract}


\end{frontmatter}

\section{Introduction}
\label{sec:intro}

Two-dimensional (2D) dilaton gravity provides a useful setting for studying the basic structure of gravitational dynamics. Although it has fewer degrees of freedom, it still captures many features of higher-dimensional gravity and often allows exact results. Because of this, it has been widely used in studies of quantum gravity~\cite{Henneaux1985,Mavromatos1989,Alwis1992}, gravitational collapse~\cite{Vaz1994,Vaz1996}, black hole physics~\cite{Bogojevic:1998ma,Nojiri:1998ww,Nojiri:1998yg,Brown1986,Russo1992,Callan1992,Bilal1993,Russo1993,EstradaJimenez2013,Frassino2015,Anacleto2016,Witten:2020ert,Ai:2020nyt,Nojiri:2020tph,Cadoni:2023tse,Barenboim2024}, thick brane world solutions~\cite{Zhong:2021gxs,Zhong2022,Zhong2024,Wang2024,Wang2024a,Yu2025}, the relativistic gravitational N-body problem~\cite{Mann:2024syg}, the AdS$_{2}$/CFT$_{1}$ correspondence~\cite{Nojiri:2022mfi}, and so on, see Refs.~\cite{Nojiri2001,Grumiller2002,Strominger:1994tn} for a review of 2D dilaton gravity and its applications.

A general 2D dilaton gravity model is defined by~\cite{Grumiller2002,Borissova:2026dilaton}
\begin{equation}
S=\frac{1}{2}\int \mathrm{d}^{2}x\,\sqrt{-g}
\left[\varphi R+U(\varphi)\mathcal{X}+W(\varphi)\right],
\label{eq:General2D}
\end{equation}
where $\varphi$ is the dilaton, $\mathcal{X}=-(\nabla\varphi)^2/2$, and $U(\varphi)$ and $W(\varphi)$ specify the theory. This class includes JT gravity~\cite{Jackiw1984,Teitelboim1983}, the CGHS model~\cite{Callan1992}, and the MMSS model~\cite{Mann1991} as special cases~\cite{Ai:2020nyt}. By a Weyl transformation, the kinetic term can be removed locally, giving the Weyl-fixed action~\cite{Banks:1990mk}
\begin{equation}
S=\frac{1}{2}\int \mathrm{d}^{2}x\,\sqrt{-g}
\left[\varphi R+W(\varphi)\right].
\label{eqai}
\end{equation}
This form is particularly useful for constructing exact black-hole solutions and for analyzing their thermodynamic charges.

Nonsingular black holes in 2D dilaton gravity have been studied in several related settings. Deformed CGHS models were shown to admit nonsingular solutions in Ref.~\cite{Bogojevic:1998ma}. Witten later considered deformations of JT gravity in the same Weyl-fixed model of Eq.~\eqref{eqai}, imposed asymptotic JT behavior, and derived the corresponding black-hole thermodynamics~\cite{Witten:2020ert}. Ai subsequently introduced a special dilaton potential giving a nonsingular black hole with a $\varphi^4$ kink-like metric function and finite curvature everywhere~\cite{Ai:2020nyt}. However, the thermodynamic quantities obtained in that construction lead to an apparent mismatch with the first law.

This apparent mismatch is part of a broader ambiguity in nonsingular black-hole thermodynamics. In many nonsingular black-hole models, the Hawking temperature, entropy-area relation, and first law cannot all be imposed simultaneously without specifying the underlying thermodynamic prescription~\cite{Mo:2006tb,Ma2014,Hennigar:2025thermodynamics}. Similar difficulties occur in higher-dimensional nonsingular black holes supported by nonlinear electrodynamics, where parameters identified with mass or charge may enter the Lagrangian itself and therefore cannot be varied as ordinary integration constants~\cite{Bardeen1968,pellicer1969,AyonBeato1998,Bronnikov:2000vy,Hayward2006,Fan:2016hvf,Cisterna:2020rkc,Barrientos:2022bzm,Li:2023yyw,Murk:2024nod,Li:2024rbw,Barrientos:2024umq,Wu:2024gqi,Karakasis:2026wes}. These examples show that the first law depends not only on local horizon data, but also on the phase space, boundary conditions, and Hamiltonian charge associated with the chosen time translation. For more details on higher-dimensional nonsingular black holes, see Refs.~\cite{Capozziello:2024atemporality,Capozziello:2025geodesics,Bueno:2024collapse,Bueno:2024puregravity,Dymnikova:2015remnants}.

In this work we use the Iyer-Wald covariant phase-space formalism~\cite{Iyer:1994ys} to derive the first law for nonsingular black holes in the Weyl-fixed theory given by Eq.~\eqref{eqai}. It will be shown that the thermodynamic phase space contains only one parameter, namely the integration constant $c$ in the metric function. The apparent violation of the first law in Ai's work arises because the energy used therein is not the Hamiltonian conjugate to the fixed time-translation generator adopted. The correct Iyer-Wald Hamiltonian energy is $E=-\frac{c}{2}$, and the first law is restored. Moreover, the energy formula agrees with the Casimir function in 2D dilaton gravity, confirming its interpretation as the physical black hole energy.

The paper is organized as follows. Section~\ref{sec:rgbh} defines the class of nonsingular black hole solutions considered in this work. Section~\ref{sec:thermo} derives the Hamiltonian energy and the first law using the Iyer-Wald formalism. Section~\ref{sec:casimir} compares the result with the Casimir function. Section~\ref{sec:conclusions} concludes.

\section{General construction of nonsingular black hole solutions}\label{sec:rgbh}
In this section, we briefly review the basic structure of 2D dilaton gravity in the Weyl-fixed frame and explain how it naturally accommodates nonsingular black hole solutions. 

For static configurations we choose the gauge
\begin{equation}\label{eqmetric1}
\mathrm{d}s^2=-A(x)\,\mathrm{d}t^2+\frac{1}{A(x)}\,\mathrm{d}x^2,
\end{equation}
where $A(x)$ is the metric function. The field equations following from Eq.~\eqref{eqai} reduce to~\cite{Witten:2020ert,Ai:2020nyt}
\begin{eqnarray}
\varphi''(x)&=&0,\label{eqai1}\\
A'(x) ~\varphi'(x)&=&W(\varphi),\label{eqai2}
\end{eqnarray}
where a prime denotes derivative with respect to $x$.
On the nonconstant-dilaton branch, Eq.~\eqref{eqai1} gives $\varphi=kx+\varphi_0$. For $k\neq0$, a shift and rescaling of the radial coordinate allow one to set~\cite{Ai:2020nyt}
\begin{equation}
\varphi=x.
\end{equation}
The curvature and the metric function are then determined directly by the dilaton potential:
\begin{equation}
R(x)=-A''(x)=-W'(x).
\end{equation}
\begin{equation}\label{eqAC}
  A(x)=\int^x W(\tilde{x})\,\mathrm{d}\tilde{x}+c=f(x)+c,
\end{equation}
where $c$ is an integration constant. Thus $W$ fixes both the theory and the profile $f(x)$, while the solution phase space associated with a fixed potential is one-dimensional and parametrized by $c$ (because there are no other parameters in the solutions for both $\varphi$ and $A(x)$). Since in two dimensions all curvature invariants are functions of $R$, a smooth potential with bounded $W'(x)$ gives a curvature-regular geometry. If $W'(x)\to0$ at the asymptotic boundary, the corresponding curvature approaches that of flat spacetime.

A black hole solution requires at least one zero of the metric function,
\begin{equation}
A(x_h)=0,
\end{equation}
which defines a Killing horizon at $x=x_h$. Near the horizon, the metric function can be expanded as
\begin{align}
A(x)&=A'(x_h)(x-x_h)+\mathcal{O}\!\left((x-x_h)^2\right),\\
    &=W(x_h)(x-x_h)+\mathcal{O}\!\left((x-x_h)^2\right).
\end{align}
For the exterior region $x>x_h$, regularity of the nonextremal Euclidean section requires~\cite{Witten:2020ert}
\begin{equation}
W(x_h)=A'(x_h) > 0.
\end{equation}
The construction of a nonsingular black hole therefore reduces to two requirements: a global regularity condition on $W'(x)$, and a horizon condition requiring a zero with $A'(x_h)> 0$. 

Ai previously obtained a nonsingular black hole from the choice
$W(\varphi)=\text{sech}^2(\varphi)$, for which the metric has the profile of a $\varphi^4$ kink
$A(x)=\tanh x+c$ ~\cite{Ai:2020nyt}. In fact, any potential
satisfying the above regularity and asymptotic conditions generates a
corresponding nonsingular geometry. For instance,
\begin{equation}
    W(\varphi)=\frac{1}{2}\text{sech}~\varphi
    \Longrightarrow 
    A(x)=\arctan(e^x)+c ,
\end{equation}
which gives a sine-Gordon-type kink profile with curvature
$R(x)=\frac{1}{2}\text{sech} x\,\tanh x$. Likewise,
\begin{equation}
    W(\varphi)=\frac{1}{1+\varphi^2}
    \Longrightarrow 
    A(x)=\arctan (x)+c ,
\end{equation}
and the scalar curvature is $R(x)=2x/(1+x^2)^2$. In both cases, the curvature is finite
for all $x$ and vanishes asymptotically, yielding regular black-hole
geometries.

\section{The first law}\label{sec:thermo}
In this section, we use the Iyer-Wald formalism to derive a thermodynamically consistent first law for the entire class of black holes defined by Eq.~\eqref{eqAC}. 

\subsection{General formalism}
Consider the Weyl-fixed dilaton gravity Lagrangian in Eq.~\eqref{eqai}, which can be written as a 2-form on the spacetime manifold as
\begin{equation}\label{eqlag}
\mathbf{L}=L \epsilon=\frac{1}{2} \left[\varphi R+W(\varphi)\right]\epsilon,
\end{equation}
where $\epsilon$ is the volume form.
Its first-order variation takes the standard form
\begin{equation}
\delta \mathbf{L} = \mathbf{E}(\psi)\,\delta\psi + d\mathbf{\Theta}(\psi,\delta\psi),
\end{equation}
where $\psi=(g_{ab},\varphi)$ denotes the collection of dynamical fields and $\mathbf{\Theta}(\psi,\delta\psi)$ is the symplectic potential. The equations of motion, given in Eqs.~\eqref{eqai1} and \eqref{eqai2}, are simply $\mathbf{E}(\psi)=0$, where $\mathbf{E}(\psi)$ is locally constructed from $\psi$ and its derivatives (see \ref{app:iw_details} for their derivation). 
For any vector field $\xi^a$, the associated Noether current is defined by~\cite{Iyer:1994ys,Wald:1993nt}
\begin{equation}
\mathbf{J}_{\xi} = \mathbf{\Theta}(\psi,\mathcal{L}_{\xi}\psi) - \xi\cdot \mathbf{L},
\end{equation}
where $\xi\cdot \mathbf{L}$ denotes the contraction of $\xi^a$ with the Lagrangian $\mathbf{L}$ and $\mathcal{L}_{\xi}\psi$ denotes the variation of the
dynamical fields induced by the diffeomorphism generated by $\xi^a$.
Diffeomorphism covariance implies~\cite{Iyer:1994ys,Wald:1993nt,Xiao2024}
\begin{equation*}
d\mathbf{J}_{\xi}=-\mathbf{E}(\psi)\,\mathcal{L}_{\xi}\psi .
\end{equation*}
Thus, on shell, the Noether current is closed for any vector field $\xi^a$, $d\mathbf{J}_{\xi}=0$, and can
therefore be written locally as
$\mathbf{J}_{\xi}=d\mathbf{Q}_{\xi}$, where $\mathbf{Q}_\xi$ is the Noether charge associated with the symmetry generated by $\xi^a$.
Further requiring $\xi^a$ to be a  Killing vector and with the property $\mathcal{L}_{\xi} \psi=0$, one obtains the following two equations~\cite{Xiao2024}:
\begin{equation}
\begin{gathered}
d \left(\delta \mathbf{Q}_{\xi}-\xi \cdot \mathbf{\Theta}(\psi,\delta \psi)\right)=0, \\
d \mathbf{Q}_{\xi}=-\xi \cdot \mathbf{L}.
\end{gathered}
\end{equation}

These two identities encode the variational and non-variational charge relations that underlie the first law and the Smarr formula. Integrating them over a Cauchy slice $\Sigma$ with boundary $\partial\Sigma$, extending from the bifurcation surface $S_H$ to a surface $S_x$ in the asymptotic region, and then applying Stokes' theorem, one obtains
\begin{align}
\int_{S_x} \left(\delta \mathbf{Q}_{\xi_H}-\xi_H \cdot \mathbf{\Theta}(\psi,\delta \psi)\right)&=\int_{S_H} \delta \mathbf{Q}_{\xi_H}, \label{eqH}\\
\int_{S_x} \mathbf{Q}_{\xi_H}-\int_{S_H} \mathbf{Q}_{\xi_H}&=-\int_{\Sigma} \xi_H \cdot \mathbf{L},\label{eqsmarr}
\end{align}
where $\xi_H$ is the horizon-generating Killing vector, normalized at the asymptotic boundary and vanishes at the bifurcation surface. In explicit calculations, one typically takes the limit $x\to\infty$. The equation~\eqref{eqsmarr} is the Iyer-Wald version of the Smarr formula.

Eq.~\eqref{eqH} is the basic Iyer-Wald statement of the first law of black hole thermodynamics, and on the right-hand side one has
$\int_{S_H}\delta\mathbf{Q}_{\xi_H}=(\kappa/2\pi)\delta S=T_H\delta S$, where $S$ is the Wald entropy formula given by~\cite{Iyer:1994ys,Wald:1993nt}
\begin{equation}\label{eqs1}
  S=\frac{2 \pi}{\kappa} \int_{S_H} \mathbf{Q}_{\xi_H},
\end{equation}
where $\kappa$ is the surface gravity defined through
\begin{equation}\label{eqkappa}
  \kappa^{2}=-\frac{1}{2}\,(\nabla_{\mu}\xi_{\nu})(\nabla^{\mu}\xi^{\nu}).
\end{equation}
For more practical calculations, Eq.~\eqref{eqs1} can be rewritten in the more explicit form~\cite{Iyer:1994ys,Wald:1993nt}
\begin{equation}\label{entropy}
S=-2 \pi \int_{S_H} E_R^{a b c d} \varepsilon_{a b} \varepsilon_{c d},
\end{equation}
where $E_R^{a b c d}=\frac{\partial L}{\partial R_{a b c d}}$ and $\varepsilon_{ab}$ is the binormal to the bifurcation surface. This is the Wald entropy formula. In Einstein gravity it reduces to the usual area law, while in more general theories it captures the corrections induced by nontrivial curvature couplings~\cite{Iyer:1994ys,Faraoni:2010yi}.

\subsection{Entropy formula}
We now use Eq.~\eqref{entropy} to derive the entropy formula for black holes in our 2D dilaton gravity model.
For the Lagrangian density in Eq.~\eqref{eqai}, we have
\begin{align}
E_R^{a b c d}&=\frac{\partial L}{\partial R_{a b c d}}=\frac14\,\varphi\left(g^{ac}g^{bd}-g^{ad}g^{bc}\right),
\end{align}
where we have used the fact that the Riemann tensor in two dimensions can be expressed in terms of the Ricci scalar as
\begin{equation}
  R_{abcd}=\frac{R}{2}\left(g_{ac}g_{bd}-g_{ad}g_{bc}\right).
\end{equation}
Now using the fact that the binormal $\varepsilon_{ab}$ is antisymmetric and normalized as $\varepsilon_{ab}\varepsilon^{ab}=-2$, one finds
\begin{equation}\label{eqs2}
S=-2\pi \int_{S_H} E_R^{a b c d} \varepsilon_{a b} \varepsilon_{c d}=2\pi \varphi_h,
\end{equation}
where $\varphi_h$ is the value of the dilaton at the horizon.

Therefore, the entropy is completely determined by the value of the dilaton at the horizon. Consequently, every black hole solution arising from the model in Eq.~\eqref{eqai} obeys the same entropy formula.

\subsection{Energy formula}
\label{subsec:energy}

We now derive the energy of the nonsingular black holes defined by Eq.~\eqref{eqAC} using the Iyer-Wald covariant phase-space formalism. The essential point is the choice of the time-translation generator. Note that due to the presence of the integration constant, the metric function may not approach unity at infinity, an asymptotically unit-normalized Killing vector would require an additional normalization factor. In the Iyer-Wald formalism, however, the Killing vector must be held fixed under phase-space variations~\cite{Wald:1993nt,Iyer:1994ys,Gao:2003ys}. We therefore fix this normalization at the outset and adopt the standard time-translation Killing vector
\begin{equation}\label{eqt}
t^a=\left(\frac{\partial}{\partial t}\right)^a,
\qquad
\delta t^a=0 .
\end{equation}

The canonical energy variation is
\begin{equation}
\delta E
=
\int_{S_{\infty}}
\left(\delta \mathbf{Q}_t-t\cdot\mathbf{\Theta}(\psi,\delta \psi)\right),
\label{eq:deltaH}
\end{equation}
which, in two dimensions, reduces to the value of the integrand at spatial infinity,
\begin{equation}
\delta E
=
\left(\delta \mathbf{Q}_t-t\cdot\mathbf{\Theta}(\psi,\delta \psi)\right)\big|_{x\to\infty}.
\label{eqenergy}
\end{equation}
For the Lagrangian in Eq.~\eqref{eqlag}, the symplectic potential and Noether charge associated with a vector field $\xi^a$ are (see \ref{app:iw_details} for the derivation)
\begin{align}
\mathbf{\Theta}_a
&=
\frac{1}{2}\epsilon_{ac}
\left[
\varphi\left(
\nabla_b\delta g^{cb}
-\nabla^c(g_{db}\delta g^{db})
\right)
-(\nabla_b\varphi)\delta g^{cb}
+(\nabla^c\varphi)g_{db}\delta g^{db}
\right],
\\
\mathbf{Q}_{\xi}
&=
-\frac{1}{2}\varphi\epsilon_{ab}\nabla^a\xi^b
-\epsilon_{ab}\xi^a\nabla^b\varphi .
\label{eq:q_general}
\end{align}
In the static gauge of Eq.~\eqref{eqmetric1}, and for the fixed generator~\eqref{eqt}, these become
\begin{equation}
\mathbf{Q}_t
=
\frac{1}{2}\varphi A'(x)-A(x)\varphi'(x),
\label{eq:q_static}
\end{equation}
and
\begin{equation}
t\cdot \mathbf{\Theta}(\psi,\delta \psi)
=
\frac{1}{2}
\left[
\varphi\,\delta A'
-\varphi'\,\delta A
\right]_{x\to\infty}.
\label{eq:tTheta_static}
\end{equation}

Since the solution phase space is one-dimensional and parametrized by the integration constant $c$ in the metric function, the variation of the fields under a change in $c$ is given by
\begin{equation}
\delta A=\delta c,
\qquad
\delta A'=0,
\qquad
\delta\varphi=0 .
\label{eq:variation_c}
\end{equation}
Using $\varphi=x$, one obtains
\begin{equation}
\delta\mathbf{Q}_t=-\delta c,
\qquad
t\cdot\mathbf{\Theta}(\psi,\delta \psi)=-\frac{1}{2}\delta c .
\label{delc}
\end{equation}
Substitution into Eq.~\eqref{eqenergy} gives
\begin{equation}
\delta E=-\frac{1}{2}\delta c .
\end{equation}
Hence the Hamiltonian energy is
\begin{equation}
E=-\frac{c}{2},
\label{eqee}
\end{equation}
which defines the energy conjugate to the fixed time-translation generator $t^a=(\partial_t)^a$. 

For the fixed generator $t^a$, the charge variation $\delta E=-\frac{1}{2}\delta c$ is an exact one-form on the
one-parameter solution branch considered here and is therefore integrable, yielding $E=-c/2$.
In a multidimensional solution phase space, by contrast, charge integrability is not automatic and must be checked explicitly,
together with the compatibility of the boundary conditions~\cite{Hajian:2015xlp,Harlow:2019yfa}.

\subsection{First law check}
\label{subsec:first_law}

We now show that the Hamiltonian energy obtained above satisfies the first law. For the static metric in Eq.~\eqref{eqmetric1}, the Hawking temperature is given by
\begin{equation}
T_H=\frac{\kappa}{2\pi}
=\frac{A'(x_h)}{4\pi}.
\end{equation}
The Wald entropy is
\begin{equation}
S=2\pi\varphi_h .
\end{equation}

The horizon position is determined by
\begin{equation}
A(x_h)=0 .
\end{equation}
For the family $A(x)=f(x)+c$, variation of this condition gives
\begin{equation}
A'(x_h)\delta x_h+\delta c=0 .
\end{equation}
Since $\varphi=x$, and for the nonextremal case, this implies
\begin{equation}
\delta\varphi_h=\delta x_h
=-\frac{\delta c}{A'(x_h)}
=-\frac{\delta c}{W(\varphi_h)} ,
\end{equation}
where we used $A'(x_h)=W(\varphi_h)$. Therefore,
\begin{equation}
T_H\delta S
=
\frac{A'(x_h)}{4\pi}
\left(2\pi\delta\varphi_h\right)
=
-\frac{1}{2}\delta c .
\end{equation}
Using the Hamiltonian variation
\begin{equation}
\delta E=-\frac{1}{2}\delta c ,
\end{equation}
we obtain
\begin{equation}
\delta E=T_H\delta S .
\end{equation}

Thus the first law holds for the static solutions of the form $A(x)=f(x)+c$ within the fixed-generator phase space specified above. In particular, Ai's nonsingular black hole belongs to this class. The apparent first-law mismatch in Ref.~\cite{Ai:2020nyt} is therefore traced not to a failure of black-hole thermodynamics, but to the use of an energy not identical to the Hamiltonian charge conjugate to the fixed time-translation Killing vector $t^a=(\partial_t)^a$.

\section{Casimir function in generic 2D dilaton gravity}
\label{sec:casimir}

It is known that generic 2D dilaton gravity theories of the form in Eq.~\eqref{eq:General2D} admit a covariantly conserved quantity, usually called the Casimir function or Casimir mass~\cite{Grumiller2002}
\begin{equation}
    C = e^{Q(\varphi)}\,Y + w(\varphi),
    \label{eq:Casimir_general}
\end{equation}
where the functions $Q(\varphi)$, $w(\varphi)$ and the scalar $Y$ are defined by
\begin{align}
    Q(\varphi) &= \int^{\varphi} U(\tilde{\varphi})\, d\tilde{\varphi}, \\
    w(\varphi) &= \frac{1}{2}\int^{\varphi} e^{Q(\tilde{\varphi})} W(\tilde{\varphi})\, d\tilde{\varphi},
    \label{eq:Qw_def}\\
    Y &= -\frac{1}{2}(\nabla \varphi)^2.
    \label{eq:Y_def}
\end{align}
For static black hole solutions, $C$ is the invariant mass parameter characterizing the geometry. In more general theories, $C$ still captures the invariant mass of the solution, but its relation to the ADM mass can be more complicated~\cite{Grumiller2002}.

We now specialize to the Weyl-fixed theory in Eq.~\eqref{eqai}.
In this case $Q(\varphi)=0$, and the conserved quantity in Eq.~\eqref{eq:Casimir_general} becomes
\begin{equation}
    C = Y + w(\varphi) = -\frac{1}{2}(\nabla\varphi)^2 + \frac{1}{2}\int^{\varphi} W(\tilde{\varphi})\,d\tilde{\varphi},
    \label{eq:Casimir_weyl_fixed}
\end{equation}
with $w'(\varphi)=\frac{1}{2}W(\varphi)$. In the static gauge of Eq.~\eqref{eqmetric1}, where $(\nabla\varphi)^2 = A(x)\,\varphi'(x)^2$, this becomes
\begin{equation}
    C = -\frac{1}{2}A(x)\,\varphi'(x)^2 + \frac{1}{2}\int^{\varphi(x)} W(\tilde{\varphi})\,d\tilde{\varphi}.
    \label{eq:Casimir_static}
\end{equation}

For the solution family in Eq.~\eqref{eqAC}, we choose the dilaton gauge $\varphi(x)=x$, so that $A'(x)=W(x)$. Eq.~\eqref{eq:Casimir_static} then reduces to
\begin{equation}
    C = -\frac{1}{2}A(x) + \frac{1}{2}\int^{x}A'(\tilde{x})\,d\tilde{x}.
    \label{eq:Casimir_equals_c}
\end{equation}
Using $A(x)=f(x)+c$ with $f'(x)=W(x)$, we obtain
\begin{equation}
    C = -\frac{1}{2}f(x) - \frac{1}{2}c + \frac{1}{2}\int^{x}f'(\tilde{x})\,d\tilde{x} = -\frac{c}{2}.
    \label{eq:Casimir_final_c}
\end{equation}
Thus, for our entire class of static solutions, the Casimir function is simply the integration constant $c$ up to the universal factor $-1/2$.

This result matches the canonical energy derived in Sec.~\ref{sec:thermo} from the Iyer-Wald formalism. In that case Eq.~\eqref{eqee} gives
\begin{equation}
  E = -\frac{c}{2}.
\end{equation}
Hence
\begin{equation}
    E=C.
\end{equation}
The agreement shows that the Casimir mass is precisely the Hamiltonian energy associated with the properly normalized asymptotic time translation.

\section{Conclusions}
\label{sec:conclusions}

We have studied the thermodynamics of nonsingular black holes in the Weyl-fixed 2D dilaton gravity model. For a fixed dilaton potential, the static solutions considered here take the form
\begin{align}
A(x)&=f(x)+c ,\\
\varphi(x)&=x ,
\end{align}
so that the thermodynamic phase space is one-dimensional and is spanned by the integration constant $c$. This simple structure makes the variational problem unambiguous once the time-translation generator is fixed.

Using the Iyer-Wald covariant phase-space formalism with the conventional fixed Killing vector
\begin{equation}
t^a=\left(\frac{\partial}{\partial t}\right)^a ,
\qquad
\delta t^a=0 ,
\end{equation}
we found that the Hamiltonian energy is
\begin{equation}
E=-\frac{c}{2}.
\end{equation}
Together with the Wald entropy $S=2\pi\varphi_h$ and the temperature $T_H=A'(x_h)/(4\pi)$, this charge satisfies
\begin{equation}
\delta E=T_H\delta S .
\end{equation}
The apparent first-law mismatch in Ai's nonsingular black hole is therefore traced to using an energy that is not the Hamiltonian charge conjugate to the fixed time-translation generator adopted in the variational principle.

We also showed that the same Hamiltonian energy coincides with the Casimir function of the corresponding 2D dilaton gravity model,
\begin{equation}
E=C=-\frac{c}{2},
\end{equation}
in this normalization convention. This identifies the Casimir charge as the conserved energy associated with the chosen asymptotic time translation. Our analysis thus gives a compact and covariant thermodynamic formulation for this class of nonsingular 2D black holes, while isolating the role of phase space and energy normalization in the first law.

\section{Acknowledgments}
This work was supported by the National Natural Science Foundation of China (Grant number 12175169).

\appendix
\section{Symplectic potential and Noether charge for our 2D dilaton gravity model}
\label{app:iw_details}

In this appendix, we derive the symplectic potential and Noether charge
used in the Iyer--Wald analysis of Sec.~\ref{sec:thermo} and clarify
their respective roles. The symplectic potential
$\mathbf{\Theta}(\psi,\delta\psi)$ is a one-form on field space and,
in the present two-dimensional theory, also a one-form on spacetime.
It arises as the boundary term in the first-order variation of the Lagrangian. It is not itself a conserved charge; rather, its antisymmetrized variation defines the
presymplectic current. More explicitly, for two commuting field
variations $\delta_1\psi$ and $\delta_2\psi$, the presymplectic current
is defined by~\cite{Iyer:1994ys,Wald:1993nt}
\begin{equation*}
\boldsymbol{\omega}
(\psi;\delta_1\psi,\delta_2\psi)
:=
\delta_1\mathbf{\Theta}(\psi,\delta_2\psi)
-\delta_2\mathbf{\Theta}(\psi,\delta_1\psi).
\end{equation*}

On the covariant phase space, the background satisfies the equations of
motion and $\delta_1\psi$ and $\delta_2\psi$ are tangent to the solution
space, so they satisfy the linearized equations of motion. Consequently,
$d\boldsymbol{\omega}=0$. For a field-independent vector field $\xi^a$, we have the identity~\cite{Iyer:1994ys,Wald:1993nt,Xiao2024}
\begin{equation*}
\boldsymbol{\omega}
(\psi;\delta\psi,\mathcal{L}_{\xi}\psi)
=
d\!\left[
\delta\mathbf{Q}_{\xi}
-\xi\cdot\mathbf{\Theta}(\psi,\delta\psi)
\right].
\end{equation*}
Now since $\mathcal{L}_{\xi}\psi=0$, the left-hand side vanishes, and the right-hand side defines a closed spacetime $0$-form in our 2D setting: $\delta\mathbf{Q}_{\xi}
-\xi\cdot\mathbf{\Theta}(\psi,\delta\psi)$, whose integral over any codimension-two  surface $S$ (which is a point in 2D) therefore defines the
Hamiltonian charge variation
\begin{equation*}
\begin{aligned}
\int_{S}
\left[
\delta\mathbf{Q}_{\xi}
-\xi \cdot\mathbf{\Theta}(\psi,\delta\psi)
\right]=\delta H_{\xi}.
\end{aligned}
\end{equation*}
If the surface is choosen on the bifurcation surface $S_H$, and by requiring $\xi^a$ to be the horizon Killing vector $\xi_{H}$, the integral gives $T_H \delta S$ as shown in Sec.~\ref{subsec:first_law}. If the surface is chosen at spatial infinity, the integral gives Hamiltonian charge variation $\delta E$ as shown in Sec.~\ref{subsec:energy}. In this way, the first law of black hole thermodynamics is established.

Consider the diffeomorphism-covariant Lagrangian $2$-form
\begin{equation}
\mathbf{L}=L\,\epsilon,
\qquad
L=\frac12\bigl(\varphi R+W(\varphi)\bigr),
\end{equation}
with independent fields $\psi=(g^{ab},\varphi)$. Its variation is
\begin{equation}
\delta \mathbf{L}
=
\frac12\,\delta\epsilon\,(\varphi R+W)
+\frac12\,\epsilon\Bigl[(R+W'(\varphi))\delta\varphi+\varphi\,\delta R\Bigr].
\end{equation}
Using
\begin{equation}
\delta\epsilon=-\frac12\,\epsilon\,g_{ab}\delta g^{ab},
\qquad
\delta R=R_{ab}\delta g^{ab}+g^{ab}\delta R_{ab},
\end{equation}
one finds
\begin{align}
\delta\mathbf{L}
&=
\frac12\,\epsilon\bigl(R+W'(\varphi)\bigr)\delta\varphi\nonumber\\
&+\frac12\,\epsilon
\Bigl[
\varphi R_{ab}-\frac12 g_{ab}(\varphi R+W)
\Bigr]\delta g^{ab}
+\frac12\,\epsilon\,\varphi\, g^{ab}\delta R_{ab}.
\label{A1}
\end{align}

Using the contracted Palatini identity
\begin{equation}
g^{ab}\delta R_{ab}=\nabla_a v^a,
\qquad
v^a=\nabla_b\delta g^{ab}-g_{cd}\nabla^a\delta g^{cd},
\label{A2}
\end{equation}
and integrating by parts, the variation takes the standard form
\begin{equation}
\delta\mathbf{L}
=
\mathbf{E}_{ab}\,\delta g^{ab}
+\mathbf{E}_{\varphi}\,\delta\varphi
+d \mathbf{\Theta}(\psi,\delta\psi),
\end{equation}
with
\begin{equation}
\mathbf{E}_{\varphi}
=
\frac12\,\epsilon\bigl(R+W'(\varphi)\bigr),
\end{equation}
and
\begin{equation}
\mathbf{E}_{ab}
=
\frac12\,\epsilon
\left[
\varphi R_{ab}
-\frac12 g_{ab}(\varphi R+W)
-\nabla_a\nabla_b\varphi
+g_{ab}\Box\varphi
\right],
\end{equation}
and symplectic potential $1$-form
\begin{eqnarray}
\mathbf{\Theta}_a(\psi,\delta\psi)&=&\frac{1}{2} \epsilon_{ac}[\varphi\left(\nabla_b \delta g^{c b}-\nabla^c\left(g_{d b} \delta g^{d b}\right)\right)\nonumber\\
&-&\left(\nabla_b \varphi\right) \delta g^{c b}+\left(\nabla^c \varphi\right) g_{d b} \delta g^{d b}],\nonumber\\
&=&\frac{1}{2} \epsilon_{ac}H^c.
\end{eqnarray}

For the boundary term appearing in the energy variation, we contract $\mathbf{\Theta}(\psi,\delta \psi)$ with the fixed asymptotic time-translation generator $t^a$. In the static gauge used here, $\det g=-1$. Choosing the orientation $\epsilon_{tx}=+1$ and setting $t^a=(\partial_t)^a$, we obtain
\begin{align}
t \cdot \mathbf{\Theta}(\psi,\delta \psi)\Big|_{x \to \infty}&=\frac{1}{2}\epsilon_{tx} H^x=\frac{1}{2} H^x,\nonumber\\
&=\frac{1}{2} [\varphi \delta A'-\varphi' \delta A]|_{x \to \infty},\nonumber\\
&=-\frac{\delta c}{2}
=\delta\!\left(-\frac{c}{2}\right),
\end{align}
which is the expression used in Eq.~\eqref{delc}.

We next derive the Noether charge. For a vector field $\xi^a$,
\begin{equation}
\mathbf{Q}^{ab}_{\xi}
=
-2E^{abcd}\nabla_c\xi_d
+
4\xi_d\nabla_c E^{abcd},
\end{equation}
where
\begin{equation}
E^{abcd}:=\frac{\partial L}{\partial R_{abcd}}.
\end{equation}
Since
\begin{equation}
R=g^{a[c}g^{d]b}R_{abcd},
\end{equation}
one has
\begin{equation}
E^{abcd}
=
\frac12\,\varphi\, g^{a[c}g^{d]b}
=
\frac14\,\varphi\left(g^{ac}g^{bd}-g^{ad}g^{bc}\right).
\label{A4}
\end{equation}
Therefore
\begin{equation}
\mathbf{Q}^{ab}_{\xi}
=
-\varphi\,\nabla^{[a}\xi^{b]}
-
2\,\xi^{[a}\nabla^{b]}\varphi.
\label{A5}
\end{equation}
The Noether charge $0$-form is thus
\begin{equation}
\mathbf{Q}_{\xi}
=
\frac12\,\epsilon_{ab}\mathbf{Q}^{ab}_{\xi}
=
-\frac12\,\epsilon_{ab}
\left(
\varphi\,\nabla^{a}\xi^{b}
+
2\,\xi^{a}\nabla^{b}\varphi
\right).
\label{A6}
\end{equation}
For the static solutions considered here, the fixed
time-translation generator used to define the energy in
Sec.~\ref{subsec:energy} also generates the Killing horizon, and
Eq.~\eqref{A6} reduces to Eq.~\eqref{eq:q_static}. With the normalization
adopted there, $\mathbf{Q}_t$ at the bifurcation surface determines
$\kappa S/(2\pi)$, whereas
$\delta\mathbf{Q}_t-t\cdot\mathbf{\Theta}(\psi,\delta\psi)$ at spatial
infinity determines the Hamiltonian energy variation $\delta E$.

In our static solution family, the potential $W(\varphi)$ fixes the solution profile, while the integration constant $c$ labels the one-dimensional solution phase
space and
determines both the horizon position and the surface gravity.
Consequently, $\mathbf{\Theta}$ and $\mathbf{Q}_t$ introduce no
additional thermodynamic variables. Instead, their endpoint
contributions relate the variation $\delta c$ to $\delta E$ at infinity
and to $T_H\delta S$ at the horizon. Thus, the thermodynamic variation
within this one-parameter family of solutions is completely captured by
the endpoint contributions entering the first law.

\bibliographystyle{elsarticle-num}






\end{document}